\documentclass[aps,preprint,showpacs]{revtex4}

\usepackage{graphicx,amssymb,amsmath,color,psfrag}
\usepackage{amsthm}
\usepackage{amsfonts}
\usepackage{algorithmic}
\usepackage{enumerate}
\usepackage{latexsym}
\usepackage{amsmath}
\usepackage{amssymb}
\usepackage{graphicx}%
\providecommand{\U}[1]{\protect\rule{.1in}{.1in}}
\begin{document}
\title{Generalization of Wolf Effect of light on Arbitrary Two-Dimensional Surface of Revolution}
\author{Chenni Xu}
\affiliation{Department of Physics, Zhejiang University, Hangzhou 310027, China}
\author{Adeel Abbas}
\affiliation{Department of Physics, Zhejiang University, Hangzhou 310027, China}
\author{Li-Gang Wang}
\email{sxwlg@yahoo.com}
\affiliation{Department of Physics, Zhejiang University, Hangzhou 310027, China}

\begin{abstract}
Investigation of physics on two-dimensional curved surface has significant
meaning in study of general relativity, inasmuch as its realizability in
experimental analogy and verification of faint gravitational effects in
laboratory. Several phenomena about dynamics of particles and electromagnetic
waves have been explored on curved surfaces. Here we consider Wolf effect, a
phenomenon of spectral shift due to the fluctuating nature of light fields, on
an arbitrary surface of revolution (SOR).
The general expression of the propagation of partially coherent beams
propagating on arbitrary SOR is derived and the corresponding evolution of
light spectrum is also obtained.
We investigate the extra influence of surface topology on spectral shift by
defining two quantities, effective propagation distance and effective
transverse distance, and compare them with longitudinal and transverse proper
lengths. Spectral shift is accelerated when the defined effective quantities
are greater than real proper lengths, and vice versa. We also employ some
typical SORs, cylindrical surfaces, conical surfaces, SORs generated by power
function and periodic peanut-shell shapes, as examples to provide concrete
analyses.
This work generalizes the research of Wolf effect to arbitrary SORs, and
provides a universal method for analyzing properties of propagation compared
with that in flat space for any SOR whose topology is known.

\end{abstract}

\maketitle

\section{\textbf{Introduction}}

A conventional and successful theory to demonstrate gravitational effect is
Einstein's general relativity (GR), which creatively relates mass to the
geometry of spacetime and provides a significantly different description of
gravitation from classical physics. However, some predictions from GR are hard
to verify due to the rather faint effects of gravitation. In recent years,
there are abundant attempts of physical systems in analogies of astronomical
phenomena, which provide ingenious platforms to simulate and explore the
theory of GR in laboratory \cite{Genov2009, Garay2000, Fedichev2003,
Leonhardt2008, Leonhardt1999, Leonhardt2000, Sheng2013, Smolyaninov2003,
Narimanov2009, Schutzhold2005, Greenleaf2007}. Innovative examples include
gravitational black holes in Bose-Einstein condensates \cite{Garay2000,
Fedichev2003}, gravitational field mimicked in moving dielectric media
\cite{Leonhardt1999, Leonhardt2000}, electromagnetic wormholes
\cite{Greenleaf2007}, etc.

Among these conceptions, one promising idea is to fabricate geometry of a
curved space-time itself with reduced dimensionality. The history of physics
on curved surface can be dated back to the early days of quantum mechanics.
Historically, two seminal works opened up the researches on curved surface
systems from different perspectives: one is to treat problem as entirely
two-dimensional (2D) from the beginning and apply canonical quantization
approach by DeWitt \cite{DeWitt1957}, another is to decouple normal and
tangent components of field vector in 3D Schr\"{o}dinger equation by
constraint procedure, therefore reducing problem into 2D by daCosta
\cite{daCosta1981}. It is believed that the curvature of space induces the
so-called geometric potential which is determined by both extrinsic and
intrinsic curvature. This potential modifies the particles' Hamiltonian and
consequently attracts much interest about dynamics of particles on surface in
condensed-matter physics \cite{Silva2015, daSilva2017}. Plenty of applications
have been presented especially since new technologies enable the synthesis of
nanostructures with complex curved geometry \cite{Fujita2005, Koshino2005,
Aoki2001, Gravesen2005, Ferrari2008, Cantele2000, Encinosa2003, Onoe2012,
Encinosa1998, Shima2009, Marchi2005, Cuoghi2009, Onoe2003}. For example, some
intriguing phenomena pertinent to electronic states, energy shifts and
electron transport have been suggested \cite{Cantele2000, Encinosa2003,
Onoe2012, Encinosa1998, Shima2009, Marchi2005, Cuoghi2009, Onoe2003}.

Dynamics of electromagnetic waves on curved surfaces was carried out in optics
about a decade ago \cite{Batz2008}. Since then light in curved space has been
investigated in various systems \cite{Batz2010, Della2010, Bekenstein2014,
Patsyk2018, Lustig2017, Spittel2015, Conti2016, Hong2017, Schultheiss2010,
Schultheiss2015, Xu2018}. For example, wave packets propagating along
nongeodesic trajectories on surfaces of revolution (SOR), demonstrating the
interaction between curvature and interference effect, is studied both
theoretically and experimentally \cite{Bekenstein2014, Patsyk2018}.
Topological phases in curved space photonics lattices are also introduced
\cite{Lustig2017}. It is demonstrated that to first order of curvature
derivative, a lattice in curved space is equivalent to that in flat space but
is subjected to an extra gauge field. Interestingly, one of curved
spaces---the SOR with constant Gaussian curvature, which corresponds to an
isotropic and uniform universe, is concretely analyzed \cite{Batz2008,
Schultheiss2010, Batz2010, Schultheiss2015}. Pioneering experimental works
have been done as well by covering a thin layer of waveguide on a 3D object.
Properties of beam propagating on such surfaces, such as evolutions of beam
width \cite{Schultheiss2010} and the second-order degree of coherence
\cite{Schultheiss2015}, are discussed in detail.

In this work, we would like to study the Wolf effect of light on arbitrary 2D
curved SORs, which is a generalization of our recent study of light on SOR
with constant Gaussian curvature \cite{Xu2018}. Wolf Effect refers to the
spectral shift of partially coherent polychromatic beams during propagation,
which was come up by Wolf in 1980s \cite{Wolf1986, Wolf1987}. This phenomenon
arises from the fluctuating (statistical) nature of light sources, and thus is
also known as correlation-induced spectral shift \cite{Wolf1996}. These years
have witnessed plenty of studies on Wolf effect in various extended areas
\cite{Leskova1997,Bocko1987,Schoonover2011,Zhu2011,Zhao2007}, for example, the
scattering system \cite{Leskova1997}, biological tissues \cite{Zhu2011},
inverse scattering problems \cite{Zhao2007}. Wolf effect has also been
experimentally verified in various systems \cite{Morris1987, Faklis1988,
Bocko1987, Schoonover2011, Zhu2011, Greffet2002, Kandpal2001, Anand2002}, such
as acoustic-correlated system \cite{Bocko1987, Schoonover2011} and ordinary
partially coherent light sources \cite{Morris1987, Faklis1988}. However,
previous works were focused in flat space. Our recent work demonstrated that
the constant Gaussian curvature of space may enhance or suppress the Wolf
effect \cite{Xu2018}.

Nevertheless, the local curvature of space, which is determined by local
distribution of mass and energy according to Einstein's field equation, is not
necessarily constant, especially in the areas around massive celestial bodies.
Therefore here we are committed to generalize our objects to research the
spectral shift of light on arbitrary SORs, and reveals acceleration or
deceleration effect of curved space on spectral shift from the perspective of
topology of surfaces.

This paper is organized as follows. In Section II, we derive point spread
function (PSF) for arbitrary SORs originating from wave equation on curved
surface, and give out the expression of output spectrum according to coherence
theory. In Section III, a theoretical analysis about the influence of surface
topology on longitudinal and transverse spectral shift is given. We develop a
technique to compare spectral shifts on such surfaces with that in flat space
by defining two quantities, effective propagation distance and effective
transverse distance. In Section IV, we employ some typical SORs, cylindrical
surfaces, conical surfaces, SORs generated by power function, and a type of
periodic peanut-shell shape to illustrate the theory mentioned above and
demonstrate some interesting phenomena. In Section V, we present our
concluding remarks.

\section{\textbf{Basic Theory}}

When being bound on 2D curved surfaces, the propagations of light beams can be
described by a 2D scalar wave equation \cite{Schultheiss2010}
\begin{equation}
\Delta_{g}\Phi+(k^{2}+H^{2}-K)\Phi=0, \label{WaveEquation basic}%
\end{equation}
where $\Delta_{g}=\partial_{i}(\sqrt{g}g^{ij}\partial_{j})/\sqrt{g}$ is the
covariant Laplacian, $g$ is determinant of metric $g_{ij}$ of curved surfaces,
and $g^{ij}=(\mathbf{g}^{-1})_{ij}$ is element of inverse matrix of
$\mathbf{g}$, $k=k_{0}n_{0}$ is the effective wave number, $k_{0}$ is the wave
number in vacuum, and $n_{0}$ is the refractive index of a thin waveguide-like
surface. This equation is derived by decoupling tangential and normal
components of 3D vectorial wave equations under neglecting polarization
effect. Here $H$ and $K$ are extrinsic and intrinsic curvature, respectively,
whose effects have been investigated in Ref. \cite{Batz2010}. The term
$H^{2}-K$, which is known as geometric potential, shows influence of curved
space on wave equation, however, in flat space both $H$ and $K$ are vanishing.
But for macroscopic cases when the scales of $H$ and $K$ are negligible
compared with $k$, only the intrinsic curvature $K$ depending on metric will
influence the light propagation through the covariant Laplacian $\Delta_{g}$.

\begin{figure}[t]
\centering
\includegraphics[width=8cm]{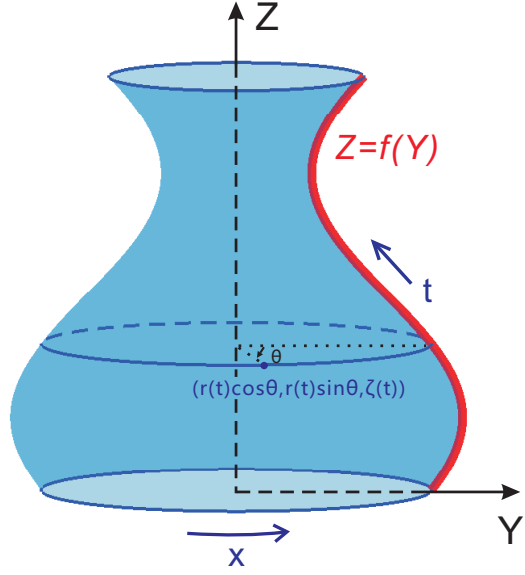}\caption{(Color online) Sketch of
surface of revolution. SOR is generated by rotating generatrix (denoted by red
solid line) in $Y-Z$ plane with respect to $Z$ axis, and $r(t)$ is radius of
rotation and $\theta$ is rotation angle. Different expression of $r(t)$ will
lead to different topology of SOR.}%
\label{Sketch of SOR}%
\end{figure}

Here we consider a sort of special surfaces --- generalized SORs, which is
formed by rotating a curve on $Y-Z$ plane with respect to one rotational axis
($Z$ axis, see Fig. \ref{Sketch of SOR}). Any SOR can be characterized by its
parametrization $r(t)$, where $r$ is known as the radius of revolution (ROR),
and $t$ is the proper length along the curve. There are two methods to
parametrize such surfaces. The first one is to directly come up with the
relation between $r$ and $t$. Another method is to first obtain the expression
of generatrix in $Y-Z$ plane, $Z=f(Y)$, and then calculate $r(t)$ according to
mathematical relation, where $f$ is a monotonic function. One can obtain the
underlying line element $ds^{2}=g_{ij}dx^{i}dx^{j}=dt^{2}+r^{2}(t)d\theta
^{2}=dt^{2}+r^{2}(t)/r_{0}^{2}d\varepsilon^{2}$ for such surfaces. Here
$r_{0}$ is the ROR at $t=0$, and $\varepsilon\equiv r_{0}\theta$ is the
transverse proper length of the initial circle of latitude defined for further
discussion. Attention should be paid that $r(t)$ is already the most general
form of metric since it can be an arbitrary form corresponding to different SORs.

With the metric $g_{ij}$, one can obtain the 2D wave equation on an SOR as
follows%
\begin{equation}
\frac{\partial^{2}\Phi(\varepsilon,t)}{\partial t^{2}}+\frac{r^{\prime}%
(t)}{r(t)}\frac{\partial\Phi(\varepsilon,t)}{\partial t}+\frac{r_{0}^{2}%
}{r^{2}(t)}\frac{\partial^{2}\Phi(\varepsilon,t)}{\partial\varepsilon^{2}%
}+k^{2}\Phi(\varepsilon,t)=0,\label{WaveEquation}%
\end{equation}
where $r\prime(t)$ is the first derivative of $r(t)$ with respect to $t$,
$H^{2}$ and $K$ are omitted when $k^{2}\gg H^{2}$, $K$. For beams starting
from $t=0$ along the longitudinal direction, substitute the ansatz $\Phi=$
$Ar^{-1/2}(t)u(t,\varepsilon)\exp\left(  ikt\right)  $ into Eq.
(\ref{WaveEquation}), where $A$ is a constant and $u(t,\varepsilon)$ is the
complex amplitude of slowly varying envelope. Thus, by applying the paraxial
approximation $\partial^{2}u(t,\varepsilon)/\partial t^{2}\ll2ik\partial
u(t,\varepsilon)/\partial t$, Eq. (\ref{WaveEquation}) evolves into%
\begin{equation}
2ik\frac{\partial u(t,\varepsilon)}{\partial t}+V_{\text{eff}}%
(t)u(t,\varepsilon)+\frac{r_{0}^{2}}{r^{2}(t)}\frac{\partial^{2}%
u(t,\varepsilon)}{\partial\varepsilon^{2}}=0,\label{SchrodingerEq1}%
\end{equation}
where
\begin{equation}
V_{\text{eff}}(t)=\frac{1}{4r^{2}(t)}\left[  \frac{dr(t)}{dt}\right]
^{2}-\frac{1}{2r(t)}\frac{d^{2}r(t)}{dt^{2}}\label{Veff}%
\end{equation}
is the effective potential intrinsically induced by the curvature of surface.
By seperating the phase $u(t,\varepsilon)=v(t,\varepsilon)\exp\left[  \frac
{i}{2k}\int\nolimits_{0}^{t}V_{\text{eff}}(t^{\prime})dt^{\prime}\right]  $
and defining a new physical quantity
\begin{equation}
\Xi(t)=\int\nolimits_{0}^{t}\frac{r_{0}^{2}}{r^{2}(t^{\prime})}dt^{\prime}.
\end{equation}
Eq. (\ref{SchrodingerEq1}) can be further simplified to%
\begin{equation}
2ik\frac{\partial v(t,\varepsilon)}{\partial\Xi}+\frac{\partial^{2}%
v(t,\varepsilon)}{\partial\varepsilon^{2}}=0,\label{SchrodingerEq2}%
\end{equation}
which is the standard 1D Schr\"{o}dinger equation. By solving Eq.
(\ref{SchrodingerEq2}), the normalized point spread function (PSF) can be
obtained%
\begin{align}
h(\varepsilon,\varepsilon^{\prime},t) &  =\sqrt{\frac{kr_{0}}{i2\pi\Xi r(t)}%
}\exp\left[  \frac{ik(\varepsilon-\varepsilon^{\prime})^{2}}{2\Xi}\right]
\nonumber\\
\text{ \ \ \ } &  \text{\ \ \ }\times\exp\left[  ikt+\frac{i}{2k}%
\int\nolimits_{0}^{t}V_{\text{eff}}(t^{\prime})dt^{\prime}\right]
.\label{PSF}%
\end{align}
Here $\varepsilon$ is the abscissa at source plane ($t=0$) and $\varepsilon
^{\prime}$ is the abscissa at any plane ($t>0$ or $t<0$) during propagation,
and $\left\{  kr_{0}/\left[  i2\pi\Xi r(t)\right]  \right\}  ^{1/2}$ is the
normalized coefficient. Before moving to next step, let us first look at the
PSF in flat space, which is usually expressed as%
\begin{align}
h_{\text{f}}(\varepsilon_{\text{f}},\varepsilon_{\text{f}}^{\prime
},z_{\text{f}}) &  =\sqrt{\frac{k}{i2\pi z_{\text{f}}}}\exp(ikz_{\text{f}%
})\nonumber\\
&  \text{ \ \ }\times\exp\left[  \frac{ik}{2z_{\text{f}}}(\varepsilon
_{\text{f}}^{2}-2\varepsilon_{\text{f}}\varepsilon_{\text{f}}^{\prime
}+\varepsilon_{\text{f}}^{\prime2})\right]  ,\label{PSF in flat}%
\end{align}
where the subscript f, denoting the flat space, is used to avoid
ambiguousness, and $z_{\text{f}}$ is the propagating distance in flat (or
free) space. Comparing Eq. (\ref{PSF}) with\textit{ }Eq.\textit{
}(\ref{PSF in flat}), one can find there is an extra phase induced by surface
in Eq. (\ref{PSF}). Besides, $z_{\text{f}}$ in Eq. (\ref{PSF in flat}) is
substituted by quantity $\Xi$ in Eq. (\ref{PSF}). Therefore, the newly-defined
quantity $\Xi$ can be regarded as the effective propagation distance, although
it intrinsically corresponds to the coordinate transformation of propagation
distance $t$.

According to coherence theory, by using PSF of Eq. (\ref{PSF}), the output
cross-spectral density at arbitrary plane $t\neq0$ can be expressed as%
\begin{align}
W_{\text{out}}(\varepsilon_{1}^{\prime},\varepsilon_{2}^{\prime},t,\omega) &
=\left\langle \Phi_{\text{out}}(\varepsilon_{1}^{\prime},t,\omega
)\Phi_{\text{out}}^{\ast}(\varepsilon_{2}^{\prime},t,\omega)\right\rangle
\nonumber\\
&  =\iint d\varepsilon_{1}d\varepsilon_{2}W_{\text{in}}(\varepsilon
_{1},\varepsilon_{2},0,\omega)\nonumber\\
&  \text{ \ \ \ \ }\times h(\varepsilon_{1},\varepsilon_{1}^{\prime}%
,t)h^{\ast}(\varepsilon_{2},\varepsilon_{2}^{\prime},t).\text{ \ \ \ \ \ \ }%
\label{OutputExpression1}%
\end{align}
Here we assume that the initial light source is a well-collimated narrow beam,
with a beam half-width $\sigma_{s}$, being incident on a macroscopic surface.
When $\sigma_{s}\ll\left\vert 2\pi r_{0}\right\vert $, the intergral interval
can be mathematically expanded to $(-\infty,+\infty)$. Without loss of
generality, the incident partially coherent source here is described by a
polychromatic Gaussian Schell-model beam
\begin{align}
W_{\text{in}}(\varepsilon_{1},\varepsilon_{2},0,\omega) &  =S_{i}(\omega
)\exp\left(  -\frac{\varepsilon_{1}^{2}+\varepsilon_{2}^{2}}{4\sigma_{s}^{2}%
}\right)  \nonumber\\
&  \text{ \ \ }\times\exp\left[  -\frac{(\varepsilon_{1}-\varepsilon_{2})^{2}%
}{2\sigma_{g}^{2}}\right]  ,
\end{align}
where $S_{i}(\omega)$ is initial spectrum, $\sigma_{s}$ is the initial beam
half-width and $\sigma_{g}$ is the initial correlation length of the source.
Correspondingly the output spectrum at any observation point can be obtained
after tedious calculation%
\begin{align}
S_{\text{out}}(x,t,\omega) &  \equiv W_{\text{out}}(\varepsilon^{\prime
},\varepsilon^{\prime},t,\omega)\nonumber\\
&  =\frac{S_{i}(\omega)}{\Omega(t,\omega)}\exp\left[  -\frac{x^{2}}%
{2\sigma_{s}^{2}\Omega^{2}(t,\omega)}\right]  ,\label{OutputSpectrum}%
\end{align}
where
\begin{equation}
\Omega(t,\omega)=\alpha\lbrack1+\Xi^{2}/Z_{R}^{2}(\omega)]^{1/2}%
\label{OMEGATW}%
\end{equation}
is the beam-expansion coefficient on arbitrary SOR, $\alpha=r(t)/r_{0}$ is the
ratio of the ROR at $t>0$ to the ROR at $t=0$ and is known as expansion ratio,
$x=\alpha\varepsilon^{\prime}$ is the proper transverse length of surface at
$t>0$, and $Z_{R}(\omega)=2k\sigma_{s}^{2}/[1+4\sigma_{s}^{2}/\sigma_{g}%
^{2}]^{1/2}=2(\omega n_{0}/c)\sigma_{s}^{2}/[1+4\sigma_{s}^{2}/\sigma_{g}%
^{2}]^{1/2}$ is the frequency-dependent Rayleigh distance of light beam. Note
that in Eq. (\ref{OutputExpression1}), the second exponential in Eq.
(\ref{PSF}) will be counteracted with its conjugate. Therefore the
curvature-induced effective potential $V_{\text{eff}}$ does not exert
influence on the output spectrum as well as spectral shift.

\section{General \textbf{behaviors of spectral shift}}

In the following discussion, the initial spectrum $S_{i}(\omega)$ of light is
assumed to be a single peak, e.g, a Gaussian or Lorentz-type spectral line,
with its spectral line-width $\delta$ being much smaller than its center
frequency $\omega_{0}$ (i.e., $\delta<<\omega_{0}$). In this sense, one can
safely suppose the output spectrum still has only a single peak with a
negligible distortion, then its new center frequency $\omega_{0}^{\prime}$ via
spectral shift must satisfy $\partial S_{\text{out}}(x,t,\omega)/\partial
\omega|_{\omega=\omega_{0}^{\prime}}=0$. From Eq. (\ref{OutputSpectrum}), one
can obtain%
\begin{equation}
\frac{S_{i}^{\prime}(\omega_{0}^{\prime})}{S_{i}(\omega_{0}^{\prime})}%
+\frac{\Omega^{\prime}(t,\omega_{0}^{\prime})}{\Omega(t,\omega_{0}^{\prime}%
)}\left[  \frac{x^{2}}{\sigma_{s}^{2}}\frac{1}{\Omega^{2}(t,\omega_{0}%
^{\prime})}-1\right]  =0, \label{Isoshift}%
\end{equation}
where $S_{i}^{\prime}(\omega_{0}^{\prime})$ and $\Omega^{\prime}(t,\omega
_{0}^{\prime})$ denote the first derivatives versus $\omega$ at $\omega
_{0}^{\prime}$. This equation has the same form as Eq. (10) in our previous
investigation \cite{Xu2018}, but the main differences are the function
$\Omega(t,\omega)$ and its derivative $\Omega^{\prime}$, which are now
generalized to any SOR. Spectral shift $\Delta\omega$ is calculated by
$\Delta\omega=\omega_{0}^{\prime}-\omega_{0}$. By solving Eq. (\ref{Isoshift}%
), in principle, one can obtain spectral shift at every spatial point on any
curved SOR, although it is too complex to give out an analytical solution.
Here we present another method by comparing the effect of curved surface on
spectral shift with the situations in flat space. Rewrite Eq. (\ref{Isoshift}) as%

\begin{equation}
\frac{x^{2}}{\alpha^{2}\sigma_{s}^{2}\Lambda^{2}(\Xi,\omega_{0}^{\prime}%
)}=1-\frac{S_{i}^{\prime}(\omega_{0}^{\prime})}{S_{i}(\omega_{0}^{\prime}%
)}\frac{\Lambda(\Xi,\omega_{0}^{\prime})}{\Lambda^{\prime}(\Xi,\omega
_{0}^{\prime})}, \label{Isoshifta}%
\end{equation}
where $\Lambda(\Xi,\omega_{0}^{\prime})=[1+\Xi^{2}/Z_{R}^{2}(\omega
_{0}^{\prime})]^{1/2}>0$, and $\Lambda^{\prime}(\Xi,\omega_{0}^{\prime}%
)=-\Xi^{2}/[\omega_{0}^{\prime}Z_{R}^{2}(\omega_{0}^{\prime})\Lambda
(\Xi,\omega_{0}^{\prime})]<0$. The correspondence in flat space is (the
detailed calculation is similar to previous one and is omitted)%
\begin{equation}
\frac{x_{\text{f}}^{2}}{\sigma_{s}^{2}\Lambda_{\text{f}}^{2}(z_{\text{f}%
},\omega_{0}^{\prime})}=1-\frac{S_{i}^{\prime}(\omega_{0}^{\prime})}%
{S_{i}(\omega_{0}^{\prime})}\frac{\Lambda_{\text{f}}(z_{\text{f}},\omega
_{0}^{\prime})}{\Lambda_{\text{f}}^{\prime}(z_{\text{f}},\omega_{0}^{\prime}%
)}, \label{Isoshiftb}%
\end{equation}
where $\Lambda_{\text{f}}(z_{\text{f}},\omega_{0}^{\prime})=[1+z_{\text{f}%
}^{2}/Z_{R}^{2}(\omega_{0}^{\prime})]^{1/2}$. From Eq. (\ref{Isoshifta}) and
Eq. (\ref{Isoshiftb}), we can find that the transverse distance $x_{\text{f}}$
on the left-hand side of Eq. (\ref{Isoshiftb}) is substituted by $x/\alpha$ on
curved surface. Actually we have known that
\begin{equation}
\varepsilon^{\prime}=x/\alpha.
\end{equation}
Thus, in this sense, $\varepsilon^{\prime}$ in Eq. (\ref{OutputSpectrum})
might be seen as the effective transverse distance, but it is not a real
length on curved surfaces. Besides, on right-hand side of Eq. (\ref{Isoshiftb}%
), the longitudinal propagation distance $z_{\text{f}}$ in flat space is
substituted by the effective propagation distance $\Xi$ of Eq.
(\ref{Isoshifta}) in curved space (defined in Section II). In the form of
$\Xi$ and $\varepsilon^{\prime}$, Eq. (\ref{Isoshifta}) is mathematically
equivalent to Eq. (\ref{Isoshiftb}). Therefore, we can find that when the
value of $\Xi>t$, compared with flat space, longitudinal spectral shift is
accelerated, and vice versa. When the value of $\varepsilon^{\prime}>x$,
transverse spectral shift is accelerated, and deceleration effect happens when
$\varepsilon^{\prime}<x$.

As is discussed in Ref. \cite{Xu2018}, longitudinally central frequency of
output spectrum moves towards higher frequency (i.e., the spectral shift is
under tendency of blue shift), while along transverse direction, central
frequency tends to move towards lower frequency (i.e., tendency of red shift).
Thus, for on-axis points where both $x$ and $\varepsilon^{\prime}$\ equal to
zero, spectral shift is free from transverse squeeze of topology of surfaces.
Since $S_{i}^{\prime}(\omega_{0}^{\prime})\Lambda(\Xi,\omega_{0}^{\prime
})=S_{i}(\omega_{0}^{\prime})\Lambda^{\prime}(\Xi,\omega_{0}^{\prime})<0$,
only blue shift occurs and increases with propagation distance. However, there
is an upper limit which can be calculated by setting $t\rightarrow\infty$ in
Eq. (\ref{Isoshift}). In general, after a long propagation, effective
propagation distance $\Xi$ is large enough so that%
\begin{equation}
\Xi^{2}/Z_{R}^{2}(\omega_{0}^{\prime})\gg1. \label{Approximation2}%
\end{equation}
Therefore, Eq. (\ref{Isoshift}) can be further simpified into%
\begin{equation}
\frac{1}{\omega_{0}^{\prime}}+\frac{S_{i}^{\prime}(\omega_{0}^{\prime})}%
{S_{i}(\omega_{0}^{\prime})}=0. \label{MaximalBS}%
\end{equation}
The solution of Eq. (\ref{MaximalBS}), which corresponds to the maximal blue
shift, is independent of topology of surface and the initial parameters of
beam, but it only depends on the incident profile of spectrum. Nevertheless,
for few SORs, the relation (\ref{Approximation2}) may not be valid, such as
the conical surface which will be mentioned in Section IV, thus Eq.
(\ref{MaximalBS}) cannot be used for these surfaces. For those exceptions, the
maximal blue shift can only be calculated by Eq. (\ref{Isoshift}), and
therefore it is determined by topology of surface, initial parameters of beams
and incident profile of spectrum.

For off-axis points where $x$ and $\varepsilon^{\prime}$\ do not equal to
zero, when $\varepsilon^{\prime2}=\sigma_{s}^{2}\Lambda^{2}(\Xi,\omega
_{0}^{\prime})$ or $x^{2}=\alpha^{2}\sigma_{s}^{2}\Lambda^{2}(\Xi,\omega
_{0}^{\prime})$, then one can obtain $S_{i}^{\prime}(\omega_{0}^{\prime})=0$,
i.e., $\omega_{0}^{\prime}=\omega_{0}$ for a single-peak spectral line. This
indicates that for the observation points along the curves of $\varepsilon
^{\prime2}=\sigma_{s}^{2}\Lambda^{2}(\Xi,\omega_{0})$ [or $x^{2}=\alpha
^{2}\sigma_{s}^{2}\Lambda^{2}(\Xi,\omega_{0})$], there is no spectral shift.
When $\varepsilon^{\prime2}<\sigma_{s}^{2}\Lambda^{2}(\Xi,\omega_{0}^{\prime
})$ or $x^{2}<\alpha^{2}\sigma_{s}^{2}\Lambda^{2}(\Xi,\omega_{0}^{\prime})$,
one can also find $S_{i}^{\prime}(\omega_{0}^{\prime})<0$ from Eq.
(\ref{Isoshifta}). Thus $\omega_{0}^{\prime}>\omega_{0}$ and it indicates blue
shift happens. When $\varepsilon^{\prime2}>\sigma_{s}^{2}\Lambda^{2}%
(\Xi,\omega_{0}^{\prime})$ or $x^{2}>\alpha^{2}\sigma_{s}^{2}\Lambda^{2}%
(\Xi,\omega_{0}^{\prime})$, one can find $S_{i}^{\prime}(\omega_{0}^{\prime
})>0$. This tells $\omega_{0}^{\prime}<\omega_{0}$ and red shift happens.

\section{\textbf{Examples of some typical SORs}}

In order to provide a concrete demonstration as well as a reasonable
verification of theories mentioned above, several examples about spectral
shift on some SORs will be given in this section. Specifically speaking, both
monotonous surfaces (i.e., the corresponding generatrices are monotonic and
hence the surfaces either contract or expand) and periodic surfaces (i.e., the
corresponding generatrices are non-monotonic and the surfaces contract and
expand periodically) are investigated. In the former case, we analyze and
proof the effect of surface by comparing longitudinal and transverse spectral
shifts on such surfaces with different power exponents. While in the latter
case some interesting phenomena occur on account of periodicity. Without loss
of generality, in the following contents initial spectral profile will be
Gaussian, i.e., $S_{i}(\omega)=\exp\left[  -(\omega-\omega_{0})^{2}%
/(2\delta^{2})\right]  /(\delta\sqrt{2\pi})$, where $\delta$\ is its line-width.

\subsection{\textbf{Cylindrical Surface}}

The ROR of a cylindrical surface is identical everywhere, i.e., $r(t)=r_{0}$
for arbitrary distance $t$, therefore the effective propagation distance is
\begin{equation}
\Xi=\int\nolimits_{0}^{t}\frac{r_{0}^{2}}{r^{2}(t^{\prime})}dt^{\prime}=t,
\end{equation}
which is exactly same as the case in flat space. Meanwhile one has always the
expansion ratio $\alpha=1$. Accordingly, beam width, behavior of longitudinal
and transverse spectral shift are all the same as that in flat space as well.
Actually the line element of a cylindrical surface is $ds^{2}=dt^{2}%
+d\varepsilon^{2}$, which is also used to describe the flat surface.

\subsection{\textbf{Conical Surface}}

Since the apex of cone is a singularity which cannot be chosen as the initial
circle of abscissa, we should first choose an initial position where the
corresponding ROR $r=r_{0}$ to establish the curvilinear coordinate system and
propagate\textit{\ }the beam (As is proved in Appendix, for beams propagating
with same trajectory on the same conical surface, however we establish the
coordinate will not influence the result). In chosen coordinate, there may be
two cases for the relations between ROR $r(t)$ and longitudinal coordinate $t$
for conical surfaces as follows%
\begin{align}
\text{Case 1}  &  \text{: }r(t)=r_{0}+\frac{1}{\sqrt{1+m^{2}}}t\ \text{for
}t\in\lbrack0,\infty)\label{rrr111}\\
\text{Case 2}  &  \text{: }r(t)=r_{0}-\frac{1}{\sqrt{1+m^{2}}}t\text{ for
}t\in\lbrack0,r_{0}\sqrt{1+m^{2}}) \label{rrr222}%
\end{align}
where $m>0$ is the absolute value of the slope of the generatrix. Now the
effective propagation distance can be calculated by%
\begin{equation}
\Xi=\int\nolimits_{0}^{t}\frac{r_{0}^{2}}{r^{2}(t^{\prime})}dt^{\prime}%
=\frac{r_{0}t}{r_{0}\pm\frac{1}{\sqrt{1+m^{2}}}t}.
\end{equation}
In the case 1, it is clearly seen that when $t\rightarrow\infty$, $\Xi$ tends
to a finite value $r_{0}\sqrt{1+m^{2}}$ although the conical SOR is expanded
to infinity. It means that its effective distance approaches a fixed value
only determined by its initial ROR\ and slope of the generatrix. Since $\Xi<t$
and $\varepsilon^{\prime}<x$, both the longitudinal and transverse spectral
shifts are decelerated. Furthermore, Eq. (\ref{MaximalBS}) mentioned in
Section III is not valid, and the maximal blue shift should be calculated by
Eq. (\ref{Isoshift}). In the case 2, as $t\rightarrow r_{0}\sqrt{1+m^{2}}$,
$\Xi$ tends to be infinity. Because of $\Xi>t$ and $\varepsilon^{\prime}>x$,
both the longitudinal and transverse spectral shifts are accelerated.

\begin{figure}[htbp]
\centering
\includegraphics[width=8cm]{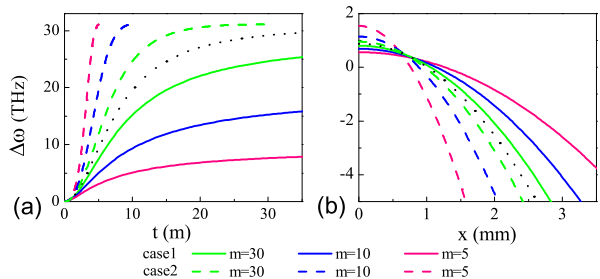}\caption{(Color online) Behaviors of
longitudinal (a) and transverse (b) spectral shifts on conical surfaces with
different slopes at (a) x=0 and (b) t=1.3 m. On each surface, propagation
starts from positions where ROR $r_{0}=1$ m. Other parameters are $\omega
_{0}=2\pi\times500$ THz, $\delta=0.1\omega_{0}$, $\sigma_{s}=1$ mm,
$\sigma_{g}=0.5$ mm, and $n_{0}=1.51$. Note that black dot lines denote the
case in flat space.}%
\label{ConicalSurface}%
\end{figure}

Figure 2 plots both longitudinal and transverse spectral shifts on conical
surfaces with different slope $m$. It is easily observed from Fig. 2 (a) that
for on-axis points, spectral shifts on surfaces of case 2 increase faster than
that in flat space, and tend to a same maximal blue shift. While on surfaces
of case 1 spectral shifts increase slower than flat space, and the
corresponding maximal blue shifts, which are always smaller than that in the
case 2, vary with slopes of conical surfaces. From Fig. 2 (b) it is also
proved that transverse spectral shifts on surfaces in the case 1 evolve
slower, and are accelerated on surfaces in the case 2. Besides, spectral shift
evolves faster with the increase of $m$ in both longitudinal and transverese
direction in the case 1, and with the decrease of $m$ in the case 2.

\subsection{\textbf{SORs generated by power function}}

Now let us consider some more complex SORs. In the $Y-Z$ plane of Fig. 1, the
expression of generatrix can be given by power function, $Z=Y^{p}-r_{0}^{p}$
\ with $Y>0$, where $r_{0}>0$ is the initial ROR at $t=0$ and $p\neq0$ is a
non-zero real number. On $Y-Z$ plane, the relation $(dY/dt)^{2}+(dZ/dt)^{2}=1$
is geometrically valid, and distance from points on generatrix to axis of
revolution ($Z$ axis), i.e., the ROR $r(t)$ constantly equals to $|Y|$. By
substituting expression of power function, relation between $r$ and $t$ is
acquired%
\begin{equation}
t=\int\nolimits_{r_{0}}^{r}\sqrt{1+p^{2}Y^{2p-2}}dY\equiv G(r),
\label{ParametrizationPF}%
\end{equation}
where $G$ is the function after integral on the right-hand side. The form of
parametrization $r(t)$ can be obtained by solving the inverse function of Eq.
(\ref{ParametrizationPF}), i.e., $r(t)=G^{-1}(t)$.

\begin{figure}[htbp]
\centering
\includegraphics[width=9cm]{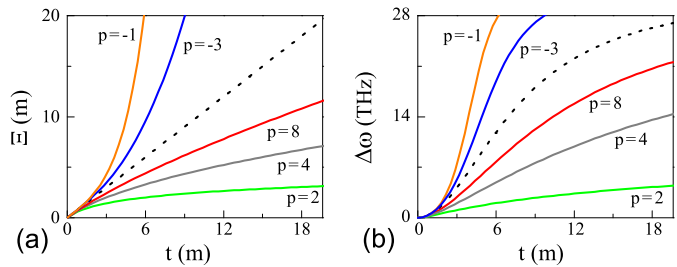}\caption{(Color online) (a)
Effective propagation distance and (b) longitudinal spectral shift versus
propagation distance $t$ for five typical SORs generated by power function
($p=-3,-1,2,4$ and $8$). For better comparison, situation of flat space is
also plotted by black dot line. Other parameters are $r_{0}=1$ m, $\omega
_{0}=2\pi\times500$ THz, $\delta=0.1\omega_{0}$, $\sigma_{s}=1$ mm,
$\sigma_{g}=0.5$ mm, $n_{0}=1.51$.}%
\label{Onaxis shift PF}%
\end{figure}

Spectral shift versus propagation distance $t$ of on-axis points on such
surfaces with different power exponent $p$\ is illustrated in Fig.
\ref{Onaxis shift PF}(b). To make the analysis more convincing, initial ROR is
chosen to be equal to $1$ m. It can be observed that on-axis longitudinal
spectral shift on surfaces with negative $p$ increases faster than that in
flat space (denoted by black dot line in figure), and it increases slower on
surfaces with positive $p$ when compared with flat space. Besides,
longitudinal spectral shift increases faster with the increase of $p$ both
when $p>0$ and $p<0$. This phenomenon can be explained as follows. For
on--axis points, since\emph{ }$\varepsilon^{\prime}=x=0$\emph{,} only
longitudinal effect of surface, to be specific, the relation between effective
propagation distance $\Xi$\ and real propagation distance $t$, should be taken
into consideration. As is shown in Fig. \ref{Onaxis shift PF}(a), effective
propagation distance $\Xi$ in flat space is smaller than that on such SORs
with negative $p$, and greater than that on such SORs with positive $p$. Both
when $p>0$ and $p<0$ effective propagation distance $\Xi$ is larger with the
increase of $p$, leading to faster spectral shift in Fig.
\ref{Onaxis shift PF}(b). The information in Fig. \ref{Onaxis shift PF}(a) can
also be obtained analytically. For SOR with $p>0$, surface is expanding, i.e.,
ROR increases over $t$. Therefore expansion ratio $\alpha$\ is greater than
$1$, and the corresponding effective propagation distance $\Xi$ is less than
its real counterpart, so longitudinal spectral shift is depressed compared
with flat space. For SOR with $p<0$, on the contrary, surface is contracting,
and effective propagation distance $\Xi$ is greater than its real counterpart,
and longitudinal spectral shift is accelerated. Besides, with greater $p$, ROR
$r(t)$ increases slower with $t$ when $p>0$ and decreases faster with $t$ when
$p<0$, which leads to smaller expansion ratio $\alpha$, and greater effective
propagation distance.

\begin{figure*}[htbp]
\centering
\includegraphics[width=13cm]{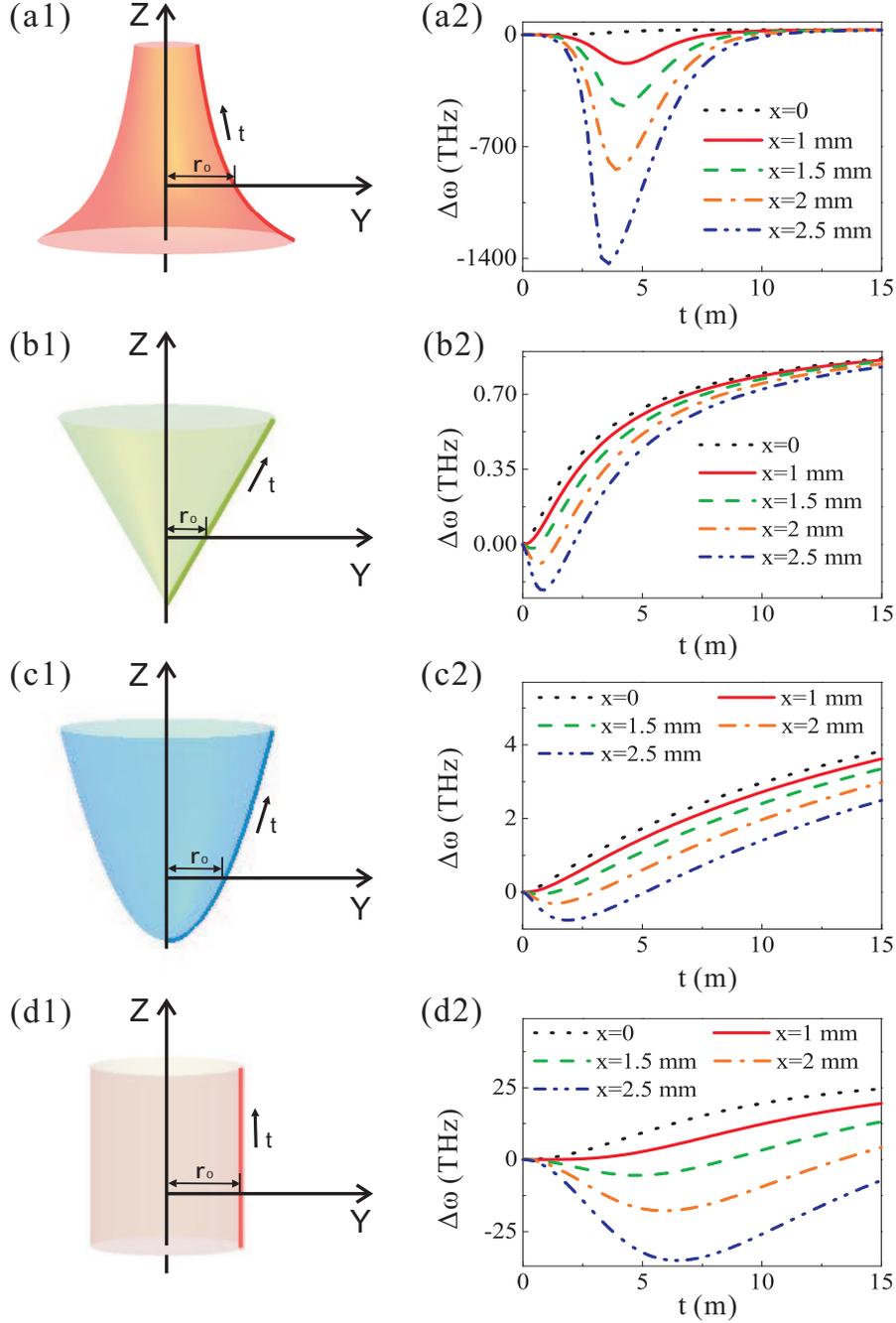}\caption{(Color online) Comparison
about distributions of spectral shifts between three typical SORs generated by
generalized power function. (a1)-(d1) Shape of such SORs with $p=-1$ (a1), $1$
(b1), $2$ (c1) together with corresponding generatrices. For better
comparison, situation on cylindrical surface, which is proved to be equivalent
to flat space, is also plotted (d1). (a2-d2) The corresponding longitudinal
spectral shifts with respect to propagation distance $t$ at different
transverse coordinates $x=0,1$ mm, $1.5$ mm, $2$ mm, $2.5$ mm on the
corresponding surfaces. Both the on-axis longitudinal spectral shift and
situation of transverse spectral shift can be analyzed from these figures. In
(a2) - (d2), other parameters are the same as Fig. \ref{Onaxis shift PF}.}%
\label{ThreePFs}%
\end{figure*}

For off-axis points, as the effective transverse distance is no longer
vanishing, its spectral shift will be subjected to both transverse and
longitudinal stretch and contraction from surface. In Fig. \ref{ThreePFs}
three representatives of such SORs ($p=-1$: Gabriel's horn, $p=1$: conical
surface, $p=2$: paraboloid) along with cylindrical surface are listed to
display the longitudinal spectral shifts at different transverse positions
($x=0$, $1$ mm, $1.5$ mm, $2$ mm, $2.5$ mm). In Figs. \ref{ThreePFs}(a2, b2,
c2, d2), it is found that in transverse direction where x coordinate
increases, spectral shift tends to develop towards red shift. Along
propagation, absolute value of red shift first increases and then decreases,
in some cases it decreases till zero and transfers to blue shift. It is
because that from its definition, effective propagation distance $\Xi$\ is a
quantity that reveals the accumulative effect of trajectory which beam passes.
At areas where longitudinal coordinate $t$ are relatively small, difference
between effective propagation distance and real propagation distance (proper
length) are not distinct, and thus longitudinal effect of surface is not
obvious and transverse effect plays dominating role. With the increase of
propagation distance, longitudinal effect of surface intensifies and spectral
shift starts to develop towards blue shift, which gives rise to a
\textquotedblleft dip\textquotedblright structure.

Comparing Figs. \ref{ThreePFs}(a2, b2, c2, d2), one may notice that spectral
(red) shift along transverse direction in Fig. \ref{ThreePFs}(a1) near the
\textquotedblleft dip\textquotedblright\ structure evolves fastest among all
the four SORs, then follow the cylindrical surface, paraboloid and conical
surface (for example, at $x=2.5$ mm, the greatest absolute value of red shift
on Gabriel' horn is approximately $1400$ THz, and approximately $30$ THz on
cylindrical surface, approximately $0.9$ THz on paraboloid, approximately
$0.2$ THz on conical surface). It is because for Gabriel' horn and actually
all such SORs with $p<0$, surface is contracting and expansion ratio $\alpha
$\ is less than $1$, transverse (red) spectral shift is drastically
accelerated. While for SORs with $p>0$, surface is expanding and expansion
ratio is greater than $1$, transverse (red) spectral shift evolves slower than
that on flat space. And with the increase of $p$, ROR increases slower with
$t$, leading to a faster transverse spectral shift.

\subsection{\textbf{Periodic peanut-shell shape (PPSS)}}

\begin{figure}[htbp]
\centering
\includegraphics[width=8cm]{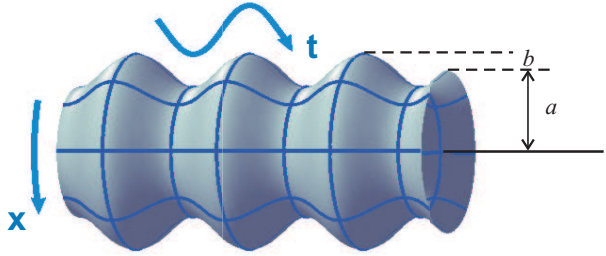}\caption{(Color online) Schematic
illustration of periodic peanut-shell shape. Curvilinear coordinates $t$ and
$x$ are marked on surface by blue lines. Parameters $a$ and $b$ denote average
radius of surface and roughness on surface, respectively.}%
\label{PeriodicShape}%
\end{figure}\ \

In this subsection, we are going to introduce a type of periodic structures on
which intriguing phenomena occur, as shown in Fig. \ref{PeriodicShape}. For
the sake of convenience, we adopt the name \textquotedblleft periodic
peanut-shell shape\textquotedblright\ from Ref. \cite{Onoe2012} to refer this
structure, which here is parametrized as%
\begin{equation}
r(t)=a-b\cos(t/R+\varphi_{0})
\end{equation}
with $a>b$. Here structure parameter $a$ determines the average radius of the
surface, while $b$ describes undulated amplitude of the surface, $0\leq
\varphi_{0}\leq\pi$ determines the initial situation of surface at $t=0$, and
$R$ denotes the longitudinal extension of the sags and crests on surface. Its
effective propagation distance is complex but can be given by%
\begin{equation}
\Xi(t)=\gamma_{1}[m\pi+\tan^{-1}\beta_{t}-\tan^{-1}\beta_{0}]-\frac
{Q\sin(t/2R)}{r(t)},
\end{equation}
where $\gamma_{1}=2aR(a-b\cos\varphi_{0})^{2}/(a^{2}-b^{2})^{3/2}$, $\beta
_{t}=(\frac{a+b}{a-b})^{1/2}\tan[(t/R+\varphi_{0})/2]$, $\beta_{0}=(\frac
{a+b}{a-b})^{1/2}\tan[\varphi_{0}/2]$, $Q=2bR(a-b\cos\varphi_{0}%
)[b\cos(t/2R)-a\cos(t/2R+\varphi_{0})]/(a^{2}-b^{2})$, and $m=\left\lfloor
\{\left\lfloor \frac{1}{\pi}(t/R+\varphi_{0})\right\rfloor +1\}/2\right\rfloor
$ with the symbol \textquotedblleft$\left\lfloor \cdot\right\rfloor
$\textquotedblright\ being a floor function. As shown in the previous
subsections, this quantity $\Xi(t)$ determines the relative spectral shift
along the on-axis propagation.

\begin{figure*}[htbp]
\centering
\includegraphics[width=13cm]{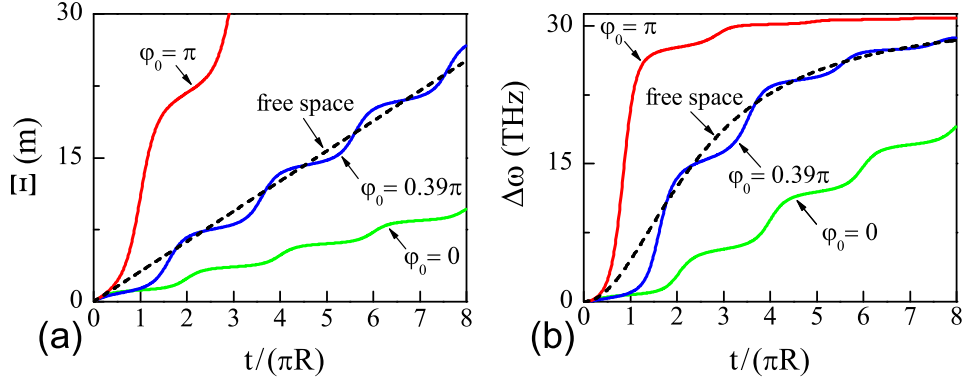}\caption{(Color online) (a) Effective
propagation distance and (b) on-axis longitudinal spectral shift of PPSSs with
three different initial phases $\varphi_{0}=0$, $0.39\pi$ and $\pi$. For
better comparison, situation in flat space is plotted in black dashed line.
Other parameters are $a=2$ m, $b=1$ m, $R=1$ m, $\omega_{0}=2\pi\times500$
THz, $\delta=0.1\omega_{0}$, $\sigma_{s}=1$ mm, $\sigma_{g}=0.5$ mm,
$n_{0}=1.51$.}%
\label{Shift PS}%
\end{figure*}

\begin{figure*}[htbp]
\centering
\includegraphics[width=16cm]{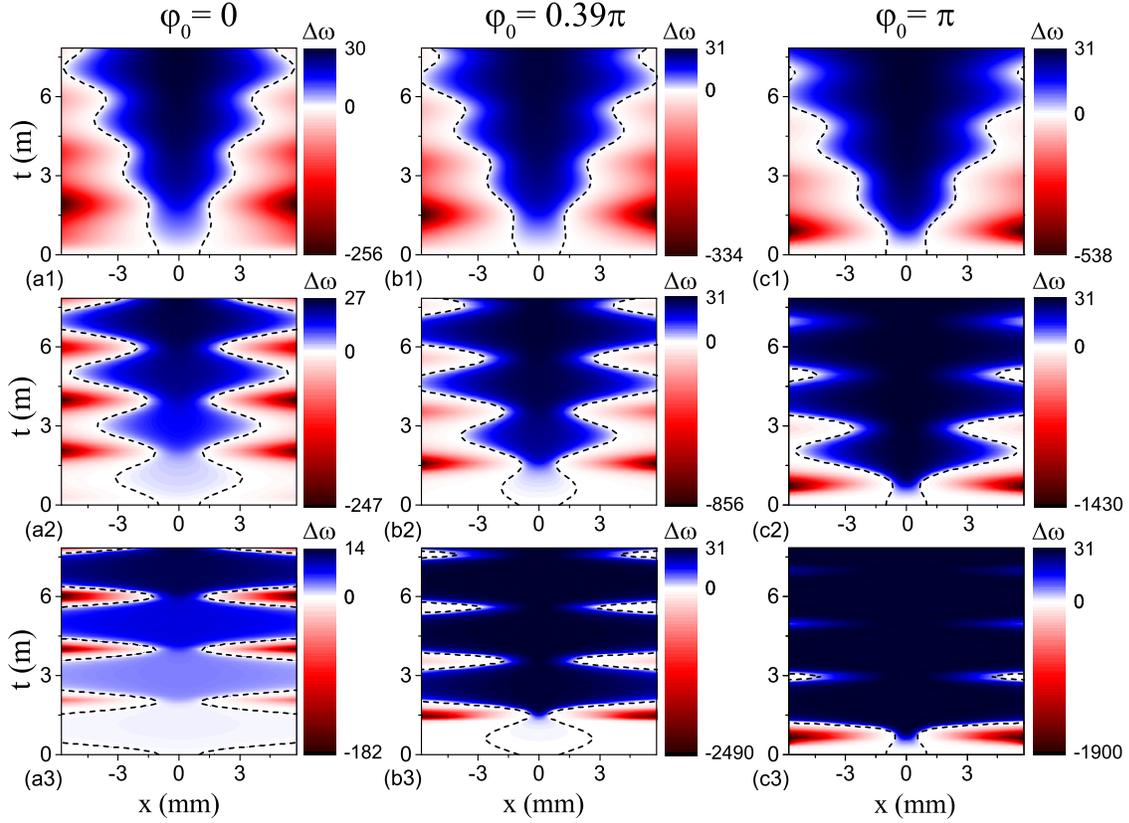}\caption{(Color online) Distribution
of spectral shift $\Delta\omega$ (THz) with different undulation on PPSSs with
three different initial phases (a1-a3) $\varphi_{0}=0$, (b1-b3) $\varphi
_{0}=0.39\pi$, (c1-c3) $\varphi_{0}=\pi$, Here $a=2$ m, $R=1.9$ m, and
undulation parameter $b$ are (a1-c1) $b=0.3$ m, (a2-c2) $b=1$ m, (a3-c3)
$b=1.7$ m. Other parameters are same as those in Fig. \ref{Shift PS}. }%
\label{DistributionPS}%
\end{figure*}

Figure \ref{Shift PS} demonstrates three typical cases of such surfaces for
different $\varphi_{0}$. The corresponding behaviors of the spectral shift
$\Delta\omega$ along the on-axis propagation are plotted in Fig.
\ref{Shift PS}(b). For example, when $\varphi_{0}=0$, one always has $\Xi<t$,
which indicates that the total spectral shift $\Delta\omega$ along the
propagating axis is smaller than that in flat space. When $\varphi_{0}%
=0.39\pi$, both the lines of $\Xi$ and $t$ may intercross together, which
leads to the effect that the total spectral shift $\Delta\omega$ along the
on-axis proagation in certain distances is larger than that in flat space and
it may also be smaller than that in flat space in other distances. When
$\varphi_{0}=\pi$, since $\Xi>t$ is always hold, spectral shift $\Delta\omega$
on the on-axis propagation in such cases is always larger than that in flat space.

Figure \ref{DistributionPS} demonstrates distributions of spectral shifts
$\Delta\omega$ on such surfaces with three different initial phases
$\varphi_{0}$, with effect of undulation on surface being considered. The
boundary between blue-shift and red-shift area (denoted by black dashed line
in each subfigure), where zero spectral shift occurs and is known as no-shift
line \cite{Xu2018}, reveals a shape of periodic expansion and contraction
during propagation, although the overall tendency is to expand, that is, the
blue-shift area tends to enlarge during propagation. By comparing Figs.
\ref{DistributionPS}(a1-a3) (or Figs. \ref{DistributionPS}(b1-b3), or Figs.
\ref{DistributionPS}(c1-c3), without loss of generality, in the following
contents we will take Figs. \ref{DistributionPS}(a1-a3) as example), it is
observed that with the increase of undulation parameter $b$ (that is, the
undulating structure on surface is more prominent in contrast to its average
radius), amplitude of oscillation of no-shift line becomes more drastic. By
comparing Figs. \ref{DistributionPS}(a1-a3), one can also observe that near
areas where contraction of no-shift line is greatest, that is, $t=2\pi R$,
$4\pi R$, $6\pi R$..., on-axis blue shift increases faster than other areas.
It is very obvious in Fig. 7(a3), where blue shift increases rapidly whenever
beam passes such positions and remains ignorable in each period. Besides, in
transverse direction, it takes shorter distance to transfer from blue-shift
area to red-shift area, and the absolute value of red shift is greater in
these areas. In conclusion, spectral shift will experience a runup at
positions where no-shift line contracts most, and this phenomenon is more
significant when undulation parameter $b$ increases. It is because that these
areas are coincidently the areas with minimal RORs, and the corresponding
expansion ratio $\alpha$\ is least. According to their definitions, effective
propagation distance $\Xi$\ increases faster, and effective transverse
distance $\varepsilon^{\prime}$ is greater, which lead to faster spectral
shift. When the amplitude of undulation increases, the contrast between the
expansion ratio at these areas $\alpha=(a-b)/(a-b\cos\varphi_{0})$ and in
other positions is more evident, resulting in more striking difference about
the speed of spectral shift at different positions. Finally, by comparing
Figs. \ref{DistributionPS}(a1, b1, c1) (or Figs. \ref{DistributionPS}(a2, b2,
c2), or Figs. \ref{DistributionPS}(a3, b3, c3)), one may also observe that
no-shift line and blue-shift area expand faster with the increase of initial
phase, and on-axis blue shift also increases faster with greater $\varphi_{0}$
(most typical in Figs. \ref{DistributionPS}(a3, b3, c3)). Since the different
initial phase $\varphi_{0}$ essentially corresponds to different relation
between $r_{0}$ and $r(t)$, and consequently leads to different evolution of
effective propagation distance $\Xi$ and effective transverse distance
$\varepsilon^{\prime}$, it also influences distribution of spectral shift,
such as blue-shift area, speed of longitudinal and transverse spectral shift, etc.

\section{Conclusion}

We have studied Wolf effect of light on arbitrary SORs. Under the paraxial
approximation, the expression of the output spectrum of polychromatic
partially coherent beams is derived by applying point spread funtion on SORs.
By defining effective propagation distance and effective transverse distance,
the effect of topology of SORs on spectral shift is comprehensively analyzed,
compared with that in flat space. If the effective propagation distance is
larger than the real correspondence, then spectral shift will be accelerated,
and vice versa. Several examples are given to verify the theoretical
calculation and analyses. This work generalizes the study of Wolf effect to
arbitrary SORs, and offers a general method to analyze effect of topology of
surface from the perspective of comparison with that in flat space.

\section*{\textbf{Appendix: Different selection of coordinates}}

In all previous analyses, beam propagation starts from origin where
longitudinal coordinate $t$ is zero. However, it is not necessary. A natural
question arises that, does spectral shift change if coordinate system is
established in different ways?

Intuitively, physics will not change by the translation of coordinate. In the
following we are going to give out detailed mathematical analysis by taking
both conditions into consideration. Let us start from derivation in Section
II. Ansatz $\Phi$ mentioned in Section II should be amended as
\begin{equation}
\Phi_{\pm}^{\text{g}}(t,\varepsilon)=A_{\pm}^{\text{g}}r^{-1/2}(t)u_{\pm
}^{\text{g}}(t,\varepsilon)\exp\left[  \pm ik(t-t_{0}^{\prime})\right]  ,
\end{equation}
where $t_{0}^{\prime}\neq0$ is the initial longitudinal coordinate, subscript
$\pm$ denotes the forward or backward direction of beam propagation. By
repeating calculation in Section II, the revised PSF can be obtained as%
\begin{align}
h_{\pm}^{\text{g}}(\varepsilon,\varepsilon^{\prime},t) &  =\sqrt{\frac
{kr_{0}^{\prime}}{i2\pi\Xi_{\pm}r(t)}}\exp\left[  \frac{ik(\varepsilon
-\varepsilon^{\prime})^{2}}{2\Xi_{\pm}}\right]  \nonumber\\
&  \text{ \ \ \ }\times\exp\left[  \pm ik(t-t_{0}^{\prime})\pm\frac{i}{2k}%
\int\nolimits_{t_{0}^{\prime}}^{t}V_{\text{eff}}(t^{\prime})dt^{\prime
}\right]  ,
\end{align}
where $r_{0}^{\prime}\equiv r(t_{0}^{\prime})$ is ROR at incident position,
and $\Xi_{\pm}=\int\nolimits_{t_{0}^{\prime}}^{t}r_{0}^{2}/r^{2}(t^{\prime
})dt^{\prime}$ is the revised effective propagation distance. The effective
potential $V_{\text{eff}}$, however, is the same as previous one, indicating
that it is independent of propagation direction and incident position.
Attention should also be paid that initial cross spectral density should be
modified as%
\begin{align}
W_{\text{in}}(\rho_{1},\rho_{2},t_{0}^{\prime},\omega) &  =S_{i}(\omega
)\exp\left(  -\frac{\rho_{1}^{2}+\rho_{2}^{2}}{4\sigma_{s}^{2}}\right)
\nonumber\\
&  \text{\ \ \ }\times\exp\left[  -\frac{(\rho_{1}-\rho_{2})^{2}}{2\sigma
_{g}^{2}}\right]  ,
\end{align}
where $\rho=r_{0}^{\prime}\theta$ is the proper length at the source plane
(i.e., incident position $t=t_{0}^{\prime})$, since at any plane except $t=0$
the coordinate $\varepsilon=r_{0}\theta$ is not the transverse proper length
as the ROR is no longer $r_{0}$. The corresponding output spectrum now is%
\begin{equation}
S_{\text{out},\pm}^{\text{g}}(x,t,\omega)=\frac{S_{i}(\omega)}{\Omega_{\pm
}^{\text{g}}(t,\omega)}\exp\left[  -\frac{x^{2}}{2\sigma_{s}^{2}\Omega_{\pm
}^{\text{g}}(t,\omega)^{2}}\right]  ,
\end{equation}
where
\begin{equation}
\Omega_{\pm}^{\text{g}}(t,\omega)=\alpha^{\prime}\left[  1+\frac{\Xi_{\pm}%
^{2}r_{0}^{\prime4}}{Z_{R}^{2}(\omega)r_{0}^{4}}\right]  ^{1/2},
\end{equation}
with new expansion ratio $\alpha^{\prime}=r(t)/r_{0}^{\prime}$. Here we can
define $\Xi_{\pm}^{\text{g}}=\Xi_{\pm}r_{0}^{\prime2}/r_{0}^{2}$, i.e.,
$\Xi_{\pm}^{\text{g}}=\int\nolimits_{t_{0}^{\prime}}^{t}r_{0}^{\prime2}%
/r^{2}(t^{\prime})dt^{\prime}$ is the generalized effective propagation
distance. Similarly, the generalized effective transverse distance becomes
$\varepsilon_{\pm}^{\prime\text{g}}=x/\alpha^{\prime}$. Obviously these two
quantities are merely relevant to ROR of incident position as well as the RORs
along the propagation path, both of which are irrelavent to how coordinate is
chosen. Therefore changing the way we establish coordinate system will not
change the calculation results of output spectrum, beam width, spectral shift,
etc. This result accords with physical intuition.

\section*{\textbf{Funding}}

Zhejiang Provincial Natural Science Foundation of China (No. LD18A040001); the
National Key Research and Development Program of China 2017YFA0304202); the
National Natural Science Foundation of China (11674284); the Fundamental
Research Funds for the Center Universities (2017FZA3005).\newline\newline

\end{document}